\documentstyle[twocolumn,aps,epsf]{revtex}
\newcommand{\tr}{\mbox{Tr} }
\newcommand{\ket}[1]{\left | #1 \right \rangle}
\newcommand{\bra}[1]{\left \langle #1 \right |}

\newcommand{\proj}[1]{\ket{#1} \! \bra{#1}}

\newcommand{\beq}{\begin{equation}}
\newcommand{\eeq}{\end{equation}}

\begin{document}
%\title{Noise enhanced quantum channel capacity}
\title{Stochastic resonance for quantum channels}
\author{
Julian Juhi-Lian Ting\footnote{Electronic address: jlting@yahoo.com}
}
\address{ No.38, Lane 93, Sec.2, Leou-Chuan E. Rd., Taichung, 40312 Taiwan, ROC}
\date{Accepted Dec. 10, 1998. Published Phys. Rev. E vol.59(3), pp2801-2803, 1999}
\maketitle
\begin{abstract}
The concept of stochastic resonance in nonlinear dynamics is applied to 
interpret the capacity of noisy quantum channels. 
The two-Pauli channel is used to illustrate the idea.
The fidelity of the channel is also considered.
Noise enhancement is found for the fidelity but not for the capacity.

\end{abstract}
%\pacs{PACS number(s): 03.67.-a, 03.67.Hk, 05.40.+j}
\section{introduction}
Stochastic resonance is a phenomenon concerned about amplifying a small 
signal forcing a nonlinear system by addition of a stochastic
noise to the signal\cite{GHJ}.
The physics behind this phenomenon is the transfer of energy from the
stochastic field into some physical process with the assistance of the
signal.

Most earlier works were concerned about periodical signals.
On the other hand, it has been pointed out by Moss in 1989 that one may
associate the switching events in a stochastic bistable and threshold system 
with an information flow through the system\cite{M}.
To consider aperiodical signals for a channel performance, 
the peak of mutual information between the input signal and output
signal is used as the definition of resonance\cite{BZ,FCB,RABI}, because 
of the informax ansatz, which uses the mutual information
to assess different ways of information processing, and because,  
traditionally, the resonance condition is defined as the peak of
the output signal to noise ratio for periodical signal cases. 

However, previous considerations for aperiodical signals
are for classical channels. 
Stochastic resonance has been studied for quantum systems\cite{GHJ}.
It is know, for periodical signal cases,
that quantum mechanics can provide additional routes by quantum 
tunneling to overcome a threshold.
Classical stochastic resonance effects can be enhanced up to two
orders of magnitude for strongly damped systems.
Therefore it will be interesting to see whether such
resonance also exits in quantum channels.

Recently, because of the development of quantum computers\cite{EJ}
people have become interested
in information transmission through noisy quantum channels\cite{S}. 
It can be used to describe processes 
such as computer memory or other secondary
storage, quantum
cryptography\cite{DEJ}, and quantum teleportation\cite{BBP}.
To study the noise enhanced channel capacities of such channels one
need a measure for noise and a measure for the channel capacities, or
a measure for any other property interested.
Therefore, the first problem need to answer is: which quantity can be
used as the {\em correct} measure for the resonance?

There are several such measures, analogous to the classical
Shannon's mutual information,
emerging during the
study of quantum computing
\cite{Schumacher96b,C,MO}. 
In this paper Schumacher's formulation of coherent information
is followed.

\section{the Noisy Channel Model}

A quantum channel  can be considered as
a process defined by an input density matrix 
$\rho_x$, and an output density matrix $\rho_y$, with the process
described by a quantum operation, ${\cal N}$,
\begin{eqnarray} 
\rho_x \stackrel{{\cal N}}{\rightarrow} \rho_y.
\end{eqnarray}
Because of decoherence, the super-operator $\cal N$
is not unitary.
However,
on a larger quantum system that includes
the environment $E$ of the system, the total evolution
operator $U_{x E}$ become unitary.  This environment may be considered
to be initially in a pure state $\ket{0_{E}}$ without
loss of generality.  In this case, the
super-operator can be written as
\begin{equation}
        {\cal N} (\rho_{x}) = \tr_{E} U_{xE} \left (
                                  \rho_{x} \otimes \proj{0_{E}} \right )
                                  {U_{xE}}^{\dagger} .
\label{channel}
\end{equation}
The partial trace $\tr_E$ is taken over environmental degree of freedom.
Eq.~(\ref{channel}) can be rewritten as
\begin{equation}
{\cal N} (\rho_x)=\sum_i A_i\rho_x A_i^\dagger\;,
\label{AnklesOfHair}
\end{equation}
in which the $A_i$ satisfy the completeness relation
\begin{equation}
\sum_i A_i^\dagger A_i = I\;,
\label{HankThoreau}
\end{equation}
which is equivalent to requiring $\tr [{\cal N} (\rho_x )]=1$.
Conversely, any set of operators $A_i$ satisfying
Eq.~(\ref{HankThoreau}) can be used in Eq.~(\ref{AnklesOfHair})
to give rise to a valid noisy channel in the sense of
Eq.~(\ref{channel}).
The mutual information in the classical formalism becomes
\begin{eqnarray}
H(x:y) = H(\rho_x)+H({\cal N}(\rho_x))-H_e(\rho_x,{\cal N}),
\end{eqnarray}
in which
$H ( \bullet ) = - \tr \left[\bullet \log_2 \bullet \right]$ 
is the von Neumann entropy\cite{N},
%$H ( {\cal N} ( \rho ) ) = - \tr {\cal N} (\rho ) \log_2 {\cal N} (\rho )$ 
and 
\begin{eqnarray}
H_e(\rho_x,{\cal N}) \equiv - \tr (W \log_2 W), \end{eqnarray}
with 
\begin{eqnarray} 
W_{ij} \equiv \mbox{Tr}(A_i \rho_x A_j^{\dagger})
\end{eqnarray}
measures the amount of information 
exchanged between the system $x$ and the environment $E$ during
their interaction\cite{S}, which 
can be used to characterize the amount of quantum noise, $N$, in the channel.
If the environment is initially in a pure state,
the entropy exchange is just the environment's entropy after the
interaction.

%The Shannon capacity $C_S$ of the classical channel is given in
%the quantum formalism by
%\begin{eqnarray}
%C_S = \max_{\rho_x} \left[ H(\rho_x) + H({\cal N}(\rho_x))-H_e(\rho_x,{\cal N})
%\right] ,
%\end{eqnarray}
%where the maximization is taken over all input states, $\rho_x$.

The coherent information is defined as
\begin{eqnarray}
C (\rho_x,{\cal N}) \equiv H \left(
        \frac{{\cal N}(\rho_x)}{\tr({\cal N}(\rho_x))} \right) -
        N (\rho_x,{\cal N}),
\end{eqnarray}
which plays a role in quantum information theory analogous to that played
by the mutual information in classical information theory.
It can be positive, negative, or zero.
Furthermore, it is     
a function of the input state and the channel only.
Although the coherence information is generally believed to
represent only a lower bound on the channel
capacity in Shannon's definition, it can be used to represent the channel 
capacity without talking about encoding\cite{BST}.

In what follows this $C-N$ 
relationship is demonstrated by the two-Pauli channel\cite{BFS}.

\section{The Two-Pauli Channel}
The {\it two-Pauli channel\/} is a noisy quantum channel on a single
qubit with
\begin{equation}
A_1=\sqrt{x}\,I\;,\;
A_2=
\sqrt{ {\scriptstyle \frac{1}{2} }
(1-x)}\,\sigma_1\;,\;
A_3=-i\sqrt{ {\scriptstyle \frac{1}{2} }
(1-x)}\,\sigma_2\;,
\end{equation}
where $I$ is the identity matrix and $\sigma_1$, $\sigma_2$, and
$\sigma_3$ are the Pauli matrices, i.e.,
\beq
\sigma_1=\left(\begin{array}{cr}
0 & 1\\
1 & 0\end{array}\right),\;\;\;\;
\sigma_2=\left(\begin{array}{cr}
0 & -i\\
i & 0\end{array}\right),\;\;\;\;
\sigma_3=\left(\begin{array}{cr}
1 & 0\\
0 & -1\end{array}\right).
\eeq
This channel has a simple interpretation:  with probability
$x$, it leaves the qubit alone; with probability $1-x$ it randomly
applies one of the two Pauli rotations to the qubit.

\begin{figure}
\epsfxsize=7.cm\epsfbox{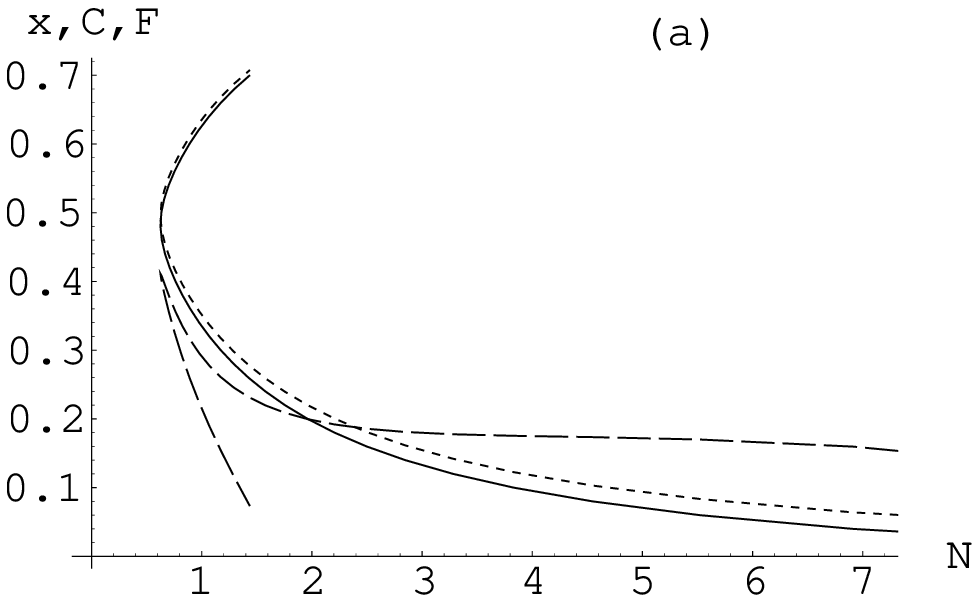}
\vskip 0.2cm
\epsfxsize=7.cm\epsfbox{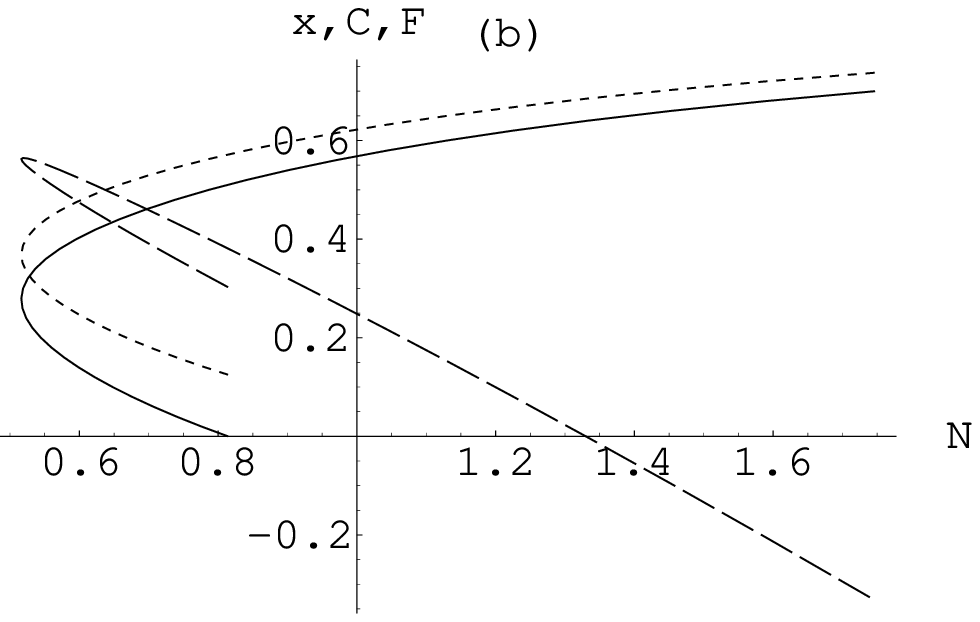}
\vskip 0.2cm
\epsfxsize=7.cm\epsfbox{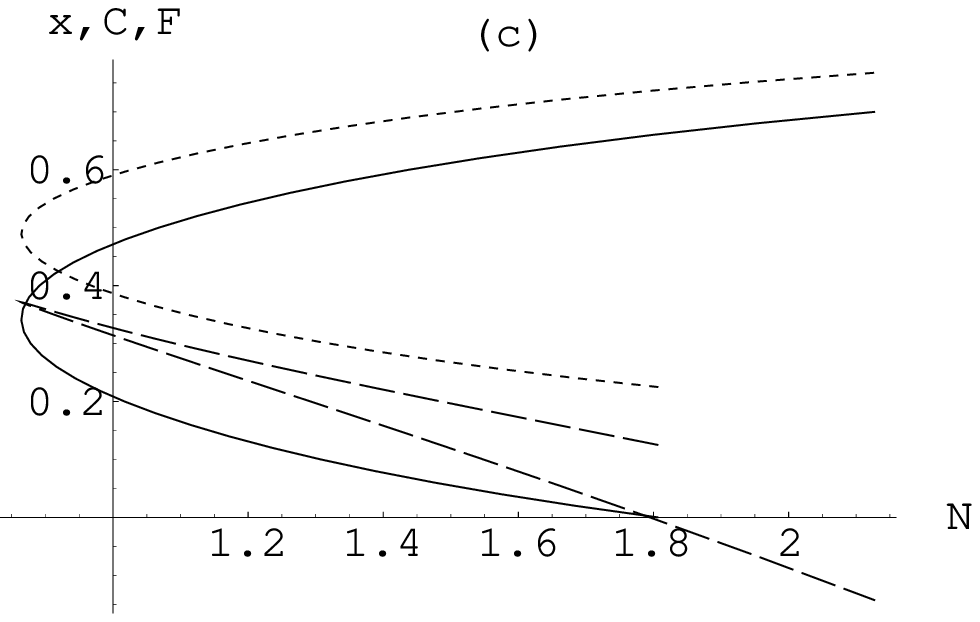}
\vskip 0.2cm
\epsfxsize=7.cm\epsfbox{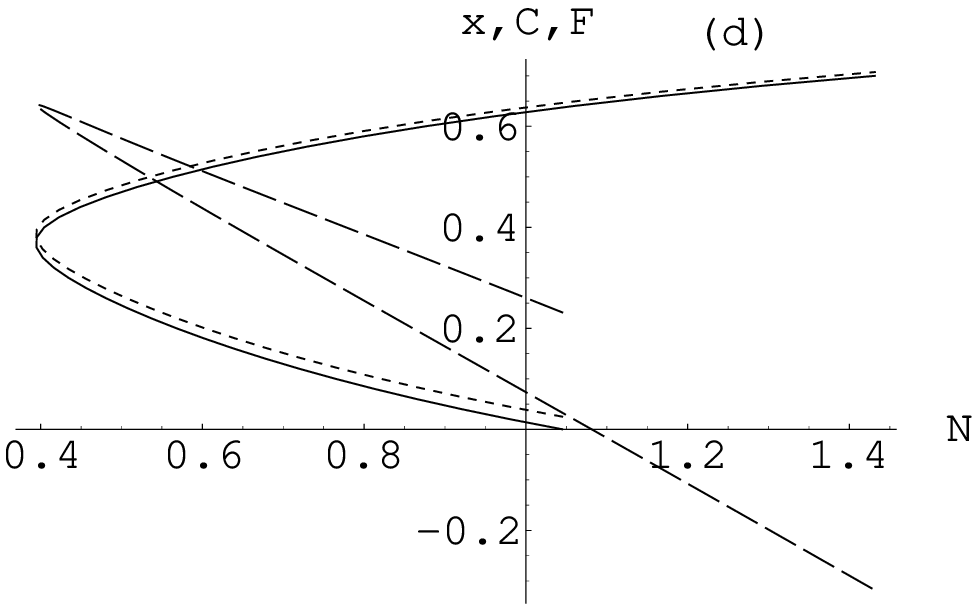}
\vskip 0.3cm
\caption{Parametric plots of $x$ versus Noise, N, (solid line);
coherence information, C, versus
noise, N, (long dashed lines);
and fidelity, F, versus noise, N, (short dashed lines) for the parameter
$x$ from $0.0$ to $0.7$ and
various initial states:
(a) $a_1 = 0.1, a_2 = 0.2, a_3 =0.9;$
(b) $a_1 = 0.3, a_2 = 0.4, a_3 =0.2;$
(c) $a_1 = 0.6, a_2 = 0.3, a_3 =0.5;$
(d) $a_1 = 0.1, a_2 = 0.2, a_3 =0.3$. 
}
\label{fig1}
\end{figure}

A general (input) state in the Bloch sphere representation
can be written as
\beq
\rho_x=\frac{1}{2}\Big(I + \vec{a}\cdot\vec{\sigma}\Big)\;.
\eeq
Here, $\vec{a}=(a_1,a_2,a_3)$ is the Bloch vector of length unity or less,
and $\vec{\sigma}$ is the vector of Pauli matrices.
For two-state systems, a
Bloch vectors with unity radius describe pure quantum states, those with
radius less than unity described mixed states, and those with
radius greater than unity do not describe any quantum state.
The action of the channel on this density matrix is:
\beq
{\cal N}(\rho_x)=
\frac{1}{2}\Big(I + \vec{b}\cdot\vec{\sigma}\Big)\;,
\eeq
in which 
\beq
\vec{b}=\Big(a_1 x,\,a_2 x,\,a_3 (2x-1)\Big)\;.
\eeq
The matrix $W$ thus computed reads,
\begin{equation}
W =\left(\begin{array}{ccc}
x & a_1 \sqrt{\frac {x (1-x)} 2} & i a_2 \sqrt{\frac { x(1-x)} 2}\\
a_1 \sqrt{{\frac {x (1-x)} 2}} & {\frac {1-x} 2} & {\frac {a_3 (1-x)} 2} \\
- i a_2 \sqrt{{\frac { x(1-x)} 2}} & {\frac {a_3 (1-x)} 2} & {\frac {1-x} 2} \\
\end{array}\right).
\end{equation}
The noise strength 
\begin{equation}
N   = - \sum_{i=1}^3 \lambda_i \log_2 \lambda_i
\end{equation}
with $\lambda_i$ been the eigenvalues of the $W$ matrix, while
\beq
H (\rho_y) = - \sum_{i=1}^2 \theta_i \log_2 \theta_i,
\eeq
with $\theta_{1,2}= [1 \pm \sqrt{(a_1^2+a_2^2) x^2+ a_3^2 (1-2x)^2}]/2$.
The noise enhancement can be investigated by looking for
some initial states, $(a_1, a_2, a_3)$, and flipping rate, $x$,
where the slope of ${\partial C}/{\partial N} >   0$.
However, in practice it is difficult to calculate such function analytically
and obtain useful results.
Some examples of the capacity-noise relation are plotted in Fig.~\ref{fig1}.
For all cases we tried we find no such enhancement. 
However, at some range
of noise the capacity is not a single valued function. With a proper choice
of the flipping rate one can indeed have higher capacity.
This is a result of the flipping rate, $x$, been a non-monotonic function
of the noise as shown in the solid lines of Fig.~\ref{fig1}.
A moderate flipping rate actually reduce the noise for all cases
plotted.
For a communication channel
the (entangled) fidelity,
\beq
F = \sum_{\mu} (\tr \rho_x A_{\mu})(\tr \rho_x A_{\mu}^{\dagger}),
\eeq
is also of our concern, since it represent the quality of the signal 
transmitted. For the two-Pauli channel
\beq
F = {\frac 1 2 }{(a_1^2 + a_2^2) ( 1-x)}  + x.
\eeq
They are plotted along with the coherent information
in Fig.~\ref{fig1}.  
The fidelity do exhibit noise enhancement clearly as shown 
in Fig.~\ref{fig1}(b)-(d).
For the best performance of a channel, there are some
trade-off between the fidelity and the capacity.
It is interesting to notice the capacity-noise curve become 
cusp and eventually
collapsed into a line as one approaches the Bloch sphere.
Note that some people might think fidelity can be used as 
a measure of the noise strength. However, it is only an {\em indirect}
measure.
It measures the effect of the noise instead of the noise itself.
As shown in the figures, the fidelity is 
close to the noise in some cases.

%As far as the present author is aware, no previous consideration of such
%capacity-noise relation has been made before, except Lloyd considered
%capacity-flipping rate relation\cite{L}.
%However, the capacity-noise relation is more physical meaningful,
%in particular, in connection with stochastic resonance.

\section{conclusion}
In summary, "quantum stochastic resonance"
in a quantum communication channel is considered in the present work.
A "resonance" of the channel capacity
is not found in the two-Pauli channel. 
On the other hand, there do have noise enhancement for fidelity.
However, 
%it has been shown by Leggett {\it et al.} that dissipation will lead to
%a lost of coherence\cite{L}.
the classical stochastic resonance phenomena is a result of interplay
between probabilistic and deterministic evolutions.
Such interplay is manifested in the coherence of quantum states.
As the dissipation enhanced quantum channel capacity is consistent with
the results found in periodical forced quantum systems,
it is very  possible for stochastic resonance for both fidelity and capacity
to exits on 
other type of sources, channels and even definition of capacity or fidelity.
There might have subtlety among different channels, as people have found
in the cases of stochastic resonances that various potential wells will
result in slightly different resonance conditions.
They are under further investigation.

%%%%%%%%%%%%%%%%%%%%%%%%%%%%%%%%%%%%%%%%%%%%%%%%%%%%%%%%%%%%%%%%%%%%%%%%%%%%%%%
%%%%%%%%%%%%%%%%%%%%%%%%%%%%%%%%%%%%%%%%%%%%%%%%%%%%%%%%%%%%%%%%%%%%%%%%%%%%%%%
\end{document}